\documentstyle[12pt]{article}
\def\be{\begin{equation}}
\def\ee{\end{equation}}
\def\ba{\begin{eqnarray}}
\def\ea{\end{eqnarray}}
\def\ga{\mathrel{\raise.3ex\hbox{$>$\kern-.75em\lower1ex\hbox{$\sim$}}}}
\def\la{\mathrel{\raise.3ex\hbox{$<$\kern-.75em\lower1ex\hbox{$\sim$}}}}

\begin{document}
\renewcommand{\thefootnote}{\fnsymbol{footnote}}
\baselineskip=21pt
\rightline{UMN--TH--1814/99}
\rightline{TPI--MINN--99/39}
\rightline{hep-ph/9909247}
\rightline{August 1999}
\vskip.25in
\begin{center}

{\large{\bf Cosmology II: From the Planck Time to
BBN}}\footnotetext{Summary of talk given at the David N. Schramm
Memorial Symposium: Inner Space/Outer Space II, Fermilab, May 1999}
\end{center}
\begin{center}
\vskip 0.5in

{Keith A. Olive}
\vskip 0.2in
{\it
{Theoretical Physics Institute, School of Physics and Astronomy, \\
University of Minnesota, Minneapolis, MN 55455, USA}}
\vskip 0.5in
{\bf Abstract}
\end{center}
\baselineskip=16pt \noindent
Progress in early Universe cosmology, including strings, extra
dimensions, inflation, phase transitions, and baryogenesis, is
reviewed.

\baselineskip=16pt

\section{Introduction}

In 1984, as astro-particle physics was beginning to thrive, the
first Inner Space/Outer Space workshop was organized \cite{isos}.
The contents of that first volume reflect well on the state of
the field at that time.  Concerning pre-big bang nucleosynthesis (BBN)
cosmology, the major focuses were centered on inflation, magnetic
monopoles, Kaluza-Klein cosmology, supersymmetry, supergravity, and
quantum gravity. Inflation was still new enough that there was a strong
interest in finding working models of inflation using realistic
particle physics. Particular emphasis was placed on inflationary
models in the context of supergravity. After the demise of
old and new inflation in the context of SU(5), the inflaton was
created to allow a framework for constructing toy models
\cite{rev}.  After fifteen years, the search for a realistic
inflationary model continues.

Magnetic monopoles were a big topic at ISOS I. The report of a
possible discovery \cite{mm} of one of these cosmological GUT relics
sparked an enormous amount of activity, which in contrast to
inflation, has largely died away. 

Beginning with SU(5) and the possibility of generating the baryon
asymmetry of the Universe, unification has been an integral part of
early Universe cosmology.  At ISOS I, while there were only a few
contributions on supersymmetry/supergravity, there was an active
session on Kaluza-Klein unification.  All of these avenues have
been incorporated (more or less) in string theory and in its parent 
(mother) theory, M-theory.

In the concluding remarks of ISOS I \cite{concl}, we made some
predictions regarding the future of the particle physics/cosmology
interface.   I quote from that paper, ``That brings us to the
future. First, on the theoretical side, it seems likely that
inflation, or at least some offshoots of the inflationary paradigm,
will continue to be promising avenues to pursue." (Given the
volumes of publications on inflation in the ensuing years, this is
at best an understatement.) ``One of the most promising approaches
to unification of all forces seems to be through additional spatial
dimensions. An area still in its infancy, cosmology with extra
dimensions adds yet another puzzling, but perhaps not unrelated,
fact to our list: why are all but three of the spatial dimensions
so small?  Superstring cosmology opens a Pandora's box of new
problems - -  whence came geometry, was there an initial
singularity, does the Universe after all have a limiting
temperature?"  It is remarkable that in the last fifteen years,
these are precisely the same questions we continue to ask.  To be
sure, much progress has been made on the technical side, but
cosmology at the Planck time remains a holy grail.

While it is not possible in the context of this contribution in
memory of Dave Schramm to completely review the state of the
cosmology from the Planck time to BBN, I will attempt to touch on
some key issues that are of particular interest today.

\section{The Planck Time}

The Planck time in cosmology is certainly the most challenging
epoch to study, as the framework in which it must be described is still
lacking.  Though there are several different approaches to this
problem, I will limit myself to that of string/M-theory \cite{lyk}.

Much of the work on string cosmology has focussed on the problem of
inflation, and I will return to that subject in the next section.
The realization that all of the different string theories, together
with 11-dimensional supergravity, can be related through dualities 
has had a major impact on how we view the Planck epoch in cosmology.
It has even led to the ultimate question as to what we mean by the
Planck epoch.   In a standard 4-dimensional model, the
gravitational action can be written as
\be
\label{staract2}
S={1 \over 2 \kappa^2}\int d^{4}x \sqrt{g} (R + \cdots)
\ee
where $\kappa^2 = 8\pi G_N = 8\pi/M_P^2$. In the context of a four
dimensional theory, there is nothing ambiguous here as the Planck
mass, determined from Newton's constant is simply $M_P = 1.2 \times
10^{19}$ GeV. Even in the context of an ``old" 10-dimensional
string theory, the 10-dimensional gravitational constant,
$\kappa_{10}^2 = 8 \pi/M_{10}^8$ is equivalent to its 4-dimensional
counterpart if the size of the 6-dimensional compact space is
Planck scale in extent, ie. $M_{10} = M_P$.
Of course, the gravitational action in string theory must be
augmented by the presence of additional fields in the gravitational
sector, most notably by the universal coupling of the string
dilaton \cite{corrs}.  Restricting our attention to only the
dilaton gravity action, we can write 
\be
\label{act10s}
S= {1 \over 2 \kappa_{10}^2} \int d^{10}x \sqrt{g_{s}}
\Bigl\{e^{-2\phi}
\Bigl(R_{s} + 4(\nabla \phi)^2 + \cdots \Bigl) \Bigl\}
\ee
Upon reduction to 4 dimensions, we recover a relatively simple
dilaton gravity system, which for a fixed dilaton (fixed by some
potential  which is perhaps induced by supersymmetry breaking), is
just Einstein gravity in 4 dimensions along with some higher
derivative curvature corrections (not shown) and additional moduli
fields (also not shown).

As noted above, M-theory is emerging as the single underlying
theory capable of unifying all particle interactions \cite{w}.
Although our understanding of M-theory is still incomplete,
its various low energy limits, and the links between them,
are known. These are the consistent string theories and 11D
supergravity, related by dualities.
A consequence of these developments
is that the dilaton
can be viewed as another modulus field in 11D
supergravity. This could have important consequences for
cosmological applications. The troubles with implementing
conventional inflationary scenarios in string theory arise because
of the dilaton and its couplings to the other modes in the string
spectrum (see below). In 11D supergravity, such
couplings are absent, and thus some of the obstacles for inflation in
dilaton-plagued string theories could
perhaps be resolved by way of M-theory.

One of the key points in the application of $M$ theory
to phenomenology is the reconciliation of the bottom-up
calculation of $M_{GUT} \sim 10^{16}$~GeV with the string
unification scale, which is close to the four-dimensional Planck
mass scale $M_4 \sim 10^{19}$~GeV. This is achieved by postulating
a large fifth dimension ${\cal R}_5 \gg M_{GUT}^{-1}$, which is
not felt by the gauge interactions, but causes the gravitational
interactions to rise with energy much faster than in the
conventional four dimensions \cite{w}. If we assume that the
compact 6-space (Calabi-Yau?) is of the size determined by the
fundamental Planck scale, $M_{11}$ so that $V_6
\sim M_{11}^{-6}$, then the fundamental Planck scale, $M_{11}$ is
related to the four dimensional Planck scale by $M_{11}^9 V_6 {\cal
R}_5 \sim M_{4}^2$. Or, 
\be
M_{11}^3 {\cal R}_5 \sim M_4^2
\ee
For ${\cal R}_5 \gg M_{11}^{-1}$, it is possible to achieve
$M_4 \gg M_{11} \sim M_{GUT}$.
In this type of scenario, one could expect
that inflation should be considered within a five-dimensional
framework.

Within this general five-dimensional framework,
two favored ranges for the magnitude of ${\cal R}_5$ can be
distinguished. One is relatively close to $M_{GUT}^{-1}$: ${\cal
R}_5^{-1} \sim 10^{12}$ to $10^{15}$~GeV, and the other could be
as low as ${\cal R}_5^{-1} \sim 1$~TeV \cite{aq}.  In this latter
case, the large dimension is not necessarily the conventional fifth
dimension of $M$ theory. Indeed, in models studied in~\cite{aq} the
large dimension may be related to what is normally considered as one
of the six ``small" dimensions that is conventionally compactified
{\it a l\`a} Calabi-Yau. Of course the physics of a very large ($\la
1$ mm) extra spatial dimension has received considerable attention
lately \cite{dim,lyk}.

As a starting point therefore, one should consider the 11D
supergravity action
\be
\label{act}
S={1 \over 2 \kappa_{11}^2}\int d^{11}x \sqrt{g} \Bigl\{R + \cdots
\Bigl\}
\ee
where  $R$ is the scalar curvature of
the 11D metric (terms involving the $3$-form potential and its
$4$-form field strength are not shown).
Reducing to 10D is done easily by assuming that the
$11^{th}$ direction is compact.  In this case, we can carry out
Kaluza-Klein reduction of (\ref{act}) to find
\begin{eqnarray}
\label{act10}
S & = & {1 \over 2 \kappa_{11}^2}\int d^{10}x \sqrt{g_{10}} {\cal
R}_{11} \Bigl\{R_{10} +
\cdots \Bigl\}
\end{eqnarray}
After a conformal rescaling $g_{10} = {\cal R}_{11}^{-1} g_s$, and
defining the dilaton by $\exp(2\phi/3) = {\cal R}_{11}$, we find
exactly the action given in eq. (\ref{act10s}).
This is precisely the effective action which describes the low
energy limit of the IIA superstring (though to be sure of the
identification, one would be required to keep track of the terms
shown here as $\cdots$). It is easy to rewrite this action in the
ten-dimensional Einstein frame, by a further conformal rescaling
$g_s = e^{\phi/2} g_E$. The action (\ref{act10s}) can be reduced
further to make contact with type IIB and heterotic theories.

If the 6-space is of the fundamental scale, then as described
above, the Universe will have passed through a phase where it
can be effectively described by a 5-dimensional space time.
A simple ansatz for the 5D metric is that of the FRW form
\be
ds^2 = - dt^2 + a^2(t) d\vec x ^2 +
 c^2(t) d \varphi^2
\label{5ans}
\ee
In this way, the dilaton expectation value $\langle \phi \rangle$
is related to the scale factor $c(t)$.  In the specific case of 
Horava-Witten type compactifications on $S^1/Z_2$, the Universe is
described by two 4D branes at the end points of the line segment.
This type of compactification is particularly well suited for 
a reduction of M-theory to heterotic string theory, where the
matter and hidden sectors sit on opposing branes. 
The cosmology of these theories has been studied at length in
\cite{low}.

\section{Inflation}

It has been known for quite
some time that it is very difficult to incorporate conventional
inflationary scenarios based on (de Sitter expansion) into the
low-energy limit of string theory \cite{nods}. The principal
obstacle has been the fact that the low-energy dynamics of the theory
contains massless scalar fields with non-minimal couplings to gravity 
whose coupling constants are precisely given by the conformal symmetry
and/or the dualities of string theory. In an expanding universe, these
fields roll during the course of the expansion as dictated by their
equations of motion, consuming the available energy and hence
decreasing the rate of expansion. That is, besides the Minkowski
space solution, there are no other solutions to Einstein's equations
(when the Gauss-Bonnet curvature squared terms are kept) of the
dilaton-gravity system with constant curvature and a stationary
dilaton. de Sitter solutions are possible if the dilaton sits in a
potential minimum (with $V \ne 0$) due to supersymmetry breaking
effects, but in this case, the theory suffers a graceful exit problem.

Instead, one typically finds solutions where
the scale factor of the universe grows as a power of time, with the
power determined by the scalar coupling constants. Once the numerical
values of these constants, fixed by string theory, are taken into
account, it has been found that the resulting power laws are too slow
to give an inflationary universe \cite{clo}. 
For example, if we ignore the compact 6-space of eq.(\ref{act10s}) and
include a cosmological term $\Lambda$
we have a 4D action
\begin{eqnarray}
\label{act4s}
S &=& {1 \over 2 \kappa_{4}^2}\int d^{4}x \sqrt{g_{s}}
\Bigl\{e^{-2\phi} \Bigl(R_{s} + 4(\nabla \phi)^2 + \cdots + \Lambda
\Bigr) +\cdots
\end{eqnarray}
where now $\kappa_4^{-2} = M_4^2/8\pi$.
In the Einstein frame this becomes
\begin{eqnarray}
\label{act4e}
S &=& {1 \over 2 \kappa_{4}^2}\int d^{4}x \sqrt{g_{s}}
\Bigl\{ \Bigl(R - 2(\nabla \phi)^2 + \cdots + e^{2\phi}\Lambda
\Bigr) + \cdots
\end{eqnarray}
In the Einstein frame, the solution to the equations of motion yield a
scale factor which grows linearly in time \cite{lindil}. However, this
expansion turns out not to be physical -- the corresponding scale
factor in the string frame is constant. To resolve this quandary, one
must compare the scale factor in each frame with some physical length
scale, such as the Compton wavelength of a massive particle.
Had we included a massive scalar field in the action (\ref{act4s}), we
would find that the scale factor relative to the Compton wavelength
$\lambda \sim m^{-1}$ is also constant. However, in the Einstein
frame, the mass term would also be dilaton dependent with mass 
$e^\phi m $, implying that the Compton wavelength also grows linearly
in time ($\phi \sim - \ln t$) and hence relative to a physical measure
of length, the Universe according to (\ref{act4e}) is also not
expanding \cite{clo}.

An interesting alternative to the standard inflationary picture in
string theory is the pre-big bang scenario \cite{pbb}.  The solutions
to the equations of motions in the string frame yield two
distinct branches often labeled ($+$) and ($-$) corresponding to
solutions which either evolve towards singularities and are
singularity-free in the past, or evolve from singularities and are
singularity-free in the future.  One of the goals of the pre-big bang
scenario is the connection of the expanding solutions in the two
branches, in such a way that the $(+)$-branch chronologically precedes
the $(-)$-branch and hence the cosmological singularity would be
removed.  It still remains to be seen if a coherent and fully
consistent description of branch-changing can be found. In fact there
are strong arguments showing the precise difficulty of a branch change
\cite{ge} though some progress has been made to resolve this graceful
exit problem
\cite{bm}.  It is interesting to note that in the string frame, both
the expanding and contracting metrics are degenerate to a single
Einstein frame metric, and that the only difference between the two
subclasses of solutions is the sign of the dilaton field. In the
context of a full 11D theory, there are several solutions which can be
found \cite{ikko} all degenerate with the known pre-big bang solutions.

There is of course no possibility in this contribution to review all
of the work on inflation in the context of string theories and
extra dimensions.  However, before moving to the subject of
baryogenesis, I will mention two possibilities in which extra
dimensions aid the implementation of inflation. For other recent work
see \cite{linf}. 

In the first \cite{ekoy}, one makes use of the higher derivative
curvature terms in the action. 
Among the first utilizations of higher-derivative curvature terms
is the Starobinsky model \cite{star}, which is based on obtaining a
self-consistent solution of Einstein's equations when they are
modified to include one-loop quantum corrections to the stress-energy
tensor $T_{\mu\nu}$. In its simplest form, the model is equivalent
to a theory of gravity with an $R^2$ correction which can be
written as \cite{star2}
\be
\label{staract}
S={1 \over 2 \kappa^2}\int d^{4}x \sqrt{g} (R + R^2/6M^2)
\ee
It is well known that this theory is
conformally equivalent to a theory of Einstein gravity plus a
scalar field \cite{whitt}.
The potential for the resulting scalar is extremely flat for
field values $\phi
\gg M_4$ and has a minimum at $\phi = 0$ with $V(\phi = 0) = 0$.
For large initial values of $\phi$, one can recognize this as an
excellent model for chaotic inflation \cite{chaotic}.

In general, quantum corrections to the right-hand side of
Einstein's equation in the absence of matter can be written as
\cite{corr}
\begin{eqnarray}
\langle T_{\mu\nu}\rangle & = & ({k_2 \over 2880 \pi^2 })(R_\mu^\rho R
_{\nu\rho}  - {2 \over 3} R R_{\mu\nu} - {1 \over 2}  g_{\mu\nu}
R^{\rho\sigma} R_{\rho\sigma} + {1 \over 4} g_{\mu\nu} R^2 )
\nonumber \\ & & + {1 \over 6}({k_3 \over 2880 \pi^2 })(2 R_{;\mu ;
\nu}    - 2g_{\mu\nu}   R^{;\rho}_{;\rho} - 2R R_{\mu\nu}  +  {1 \over
2} g_{\mu\nu} R^2 )
\label{tmncor}
\end{eqnarray}
where $k_2$ and $k_3$ are constants that appear in the process of
regularization. $k_2$ is related to the number of
light spin states, which can be very large in variants of string
theories based on $M$ theory. On the
other hand, the coefficient $k_3$ is independent of the number of
light states. This term is equivalent to the variation of the
$R^2$ term in the effective action. The theory admits a de Sitter
solution which can be found from the $00$ component of
gravitational equation of motion \cite{dc}. Defining $H' =
2880\pi^2
/ k_2$  and $M^2  = 2880\pi^2 / k_3$, and setting the spatial curvature
$k=0$, one finds \cite{vil}
\be
H^2(H^2 - H'^2) = (H'^2/M^2)(2{\ddot H}H + 2H^2{\dot H} - {\dot H}^2 )
\label{h'}
\ee
where $H$ is the Hubble parameter. The de Sitter
solution corresponds to $H = H'$ and of course ${\dot H} = {\ddot
H} = 0$.

In order to avoid the overproduction
of gravitons there is a {\em lower} limit on the parameters
$k_{2,3}$ \cite{fp,vil}: $k_2 \ga 10^{10}$ implying the need for
billions of spin degrees of freedom to be present. While this seems
like an inordinately
large number, it is possible to generate very large
numbers of degrees of freedom in theories with extra dimensions. 
Particularly if
$N \sim M_{GUT} {\cal R}_5 \ga 10^8$ \cite{ekoy}.
The bound for $k_3$ is $k_3 \ga  10^9$,
corresponding to $M \la 10^{14}$ GeV.

In general $R^2$ corrections to the action do not appear in 5 (11)
dimensions.  Therefore the first correction is of order $R^4$.
Like the curvature squared correction, the action
\be
 S= \int d^{5}x \sqrt{G_5} \Bigl\{{M_5^3 \over 16 \pi}
R_5 + \alpha M_5^{-3} R^4_5 \Bigr\} \label{f5d}
 \ee
can be conformally transformed to a 5D Einstein theory with an
additional scalar field ($\chi$) and a potential $U(\chi)$
\cite{Maeda}. Making the KK reduction to 4D, we find in addition to
$\chi$, a second scalar field $(\phi)$ which is the modulus of the 
5th dimension and is related to the dilaton.
The resulting
potential takes the form \cite{ekoy}
\be \label{potential}
U(\phi,\chi)\sim M_4^2
M_5^2
e^{-\sqrt{\frac{2}{3}} \kappa \phi} e^{-\frac{5 \sqrt{3}}{6}
\kappa \chi} \bigl( e^{\frac{\sqrt{3}}{2} \kappa
\chi}-1\bigr)^{\frac{4}{3}}
\ee

The dilaton potential here, as in most string descendent models, is
problematic unless it is fixed by an additional potential.  In the
absence of a dilaton potential, this model will not inflate.  The
remaining potential (of $\chi$) differs from that in the $R^2$
model.  As one can see it is no longer flat, but rather takes the
shape similar to that of the double well potential. As a result,
chaotic inflation
\cite{chaotic} with a large initial value of $\chi$ is impossible
here. Nevertheless there may remain the possibility to realize
inflationary expansion in this model by using the potential energy
around the local maximum,
$V(\chi_m)$, as in the topological inflation scenario of Linde and
Vilenkin \cite{topological}. In this scenario, if the scalar field
$\chi(x)$ is randomly distributed initially with a large
dispersion, some part of the universe will roll to $\chi=0$, while
in other parts it will run away to infinity. Between any two such
regions there will appear domain walls, containing a large energy
density, $\rho \sim V(\chi_m)$. If the wall is thicker than the
Hubble radius of this energy density, there will exist a
sufficiently large quasi-homogeneous region, filled with large
potential energy, where inflationary expansion naturally sets in.

It can be shown \cite{ekoy} that the potential (\ref{potential})
does satisfy the conditions for topological inflation.
In addition, there is no problem with the graceful exit.
However, the spectrum of density fluctuations can be calculated 
and the spectral index is given by
$n_s = 1 - m^2/H_m^2$, where $m^2$ and $H_m^2$ are the curvature
(of the potential) and Hubble parameter determined at the maximum.
In this model, $n_s = 3/8$. Furthermore, the magnitude of density
fluctuations requires $H_m \sim M_5 \sim 10^{-14} M_4$ which are
probably unacceptably small.

These problems can be remedied by either the inclusion of an $R^2$
correction to the action, or by a large tower of KK states.
The resolution of the problem can be traced to the magnitude of the
Hubble parameter, which is determined by $H^2 \sim U(\chi)/M_4^2$.
In the presence of a large number of degrees of freedom, quantum
corrections modify this relation along the line of eq.
(\ref{h'}) in which case $H$ is driven to $H'$ and need not be
exceptionally small.  It suffices now that $H,M_5 \sim 10^{-5} M_4$
with ${\cal R}_5^{-1} \sim O(100)$ GeV.  In this case, the spectral
index is $\simeq 0.95$ and the magnitude of fluctuations is 
acceptably small.

The second possibility \cite{yko1,yko2}, also makes use of the
tower of KK states.  The idea of using multiple fields to drive
inflation where the parameters of the theory would normally not 
lead to sufficient inflation is called assisted inflation
\cite{liddle}.  For example, it is well known that scalar fields
with exponential potentials of the form $V(\phi) = e^{-\lambda
\phi}$, lead to power law expansion with the cosmological scale
factor growing as 
$R(t) \sim t^{p}$ and $p = 2/\lambda^2$. Density
fluctuations are no longer scale invariant but scale as 
$|{\delta \rho \over \rho}(k)|^2 \sim k^{n-1}$ with $n= 1 - {2 \over
p-1}$. Sufficient inflation along with $n \simeq 1$, requires
$p$ to be large.  In \cite{liddle}, it was shown that
a system of scalar fields each with exponential
potentials, (even if the individual powers, $p_i$, are not sufficiently
large to generate inflation) has an attractor
solution in which the universe power-law expands with a power given by
$p = \sum p_i$.

Assistance, can in fact be generalized to other types of potentials
\cite{yko1}. For example, we can
consider a general field theory of multiple, self-interacting
scalar fields of the form
\be
-{\cal L}= \sum_{i=1}^N \Biggl\{ 
\frac 12\,(\partial \phi_i)^2+  \frac{m^2}{2}\,\phi_i^2 \Biggr\} + 
\sum_{i=1}^N \Biggl\{\frac{\lambda_3}{3!}\,\phi_i^3 +
\frac{\lambda_4}{4!}\,\phi_i^4 \Biggl\}\,\,.
\ee
This system consists
of $N$ completely equivalent, decoupled scalar fields. As a result, the
Lagrangian can be written as
\ba
-{\cal L} &=& N\,\Biggl\{\frac 12\,(\partial \phi_1)^2 +
\frac{m^2}{2}\,\phi_1^2 + \frac{\lambda_3}{3!}\,\phi_1^3 +
\frac{\lambda_4}{4!}\,\phi_1^4 \Biggr\}
\nonumber \\[2mm]
&=& \frac 12\,(\partial \tilde{\phi})^2 +
\frac{m^2}{2}\,\tilde{\phi}^2 +
\frac{\tilde{\lambda}_3}{3!}\,\tilde{\phi}^3 +
\frac{\tilde{\lambda}_4}{4!}\,\tilde{\phi}^4 \,\,,
\ea
where
\be
\tilde{\phi}=\sqrt{N}\,\phi_1 \quad , \quad
\tilde{\lambda}_3= \frac{\lambda_3}{\sqrt{N}} \quad, \quad
\tilde{\lambda}_4= \frac{\lambda_4}{N} \,\,.
\label{scale}
\ee
 The resulting theory describes a single
scalar field with the same type of self-inter\-actions compared to the
fields in the original theory. However, these self-interactions are
considerably weaker since both of the coupling constants now scale
with the number of scalar fields
$N$. As a result, as the number of scalar fields that we include in the
theory becomes larger, the coupling constants become smaller and the
corresponding fine-tuning becomes milder.
Thus the same basic idea expounded in \cite{liddle,mw} carries over very
simply to chaotic inflation \cite{chaotic} based on a quartic
potential. While
$\tilde
\lambda_4$ must still be of order $10^{-12}$, the fundamental coupling in
the theory $\lambda_4$ can now be much larger if $N$ is large. 
In addition, the problem associated with chaotic inflation
\cite{em}, large ($\phi \gg M_p$) initial field values and the
necessary fine tuning of non-renormalizable interactions, is cured
\cite{yko2}.  Because of the rescaling in eq. (\ref{scale}), in a
theory of multiple fields, chaotic inflation will operate, even
though none of the fields have expectation values greater than
$M_P$.

In \cite{yko1,yko2}, it was suggested that the source of the
multiple fields is the KK reduction from a higher dimensional
theory to 4D.  For example, the reduction of a 5D theory results in 
$N \simeq {\cal R}_5 M_5$ fields.  Though these fields are not
decoupled as in the example above, and hence assistance is not
guaranteed \cite{yko1,maz}, attractor solutions do exist and
assistance is generated primarily due to the rescaling of the
fields (from 5D to 4D) which is necessary due to the dimensional
reduction. To see this, consider the action
\be
S_\phi= -\int d^{5}x \sqrt{G_5} \Bigl\{{G_5}^{AB} \partial_A \hat \phi
\,\partial_B \hat \phi +
\frac{\hat \lambda}{4!M_5}\,\hat \phi^4\,\,\Bigl\}
\ee
When KK-reduced to 4D, $\sqrt{G_5} \to \sqrt{G_4} {\cal R}_5 $
and canonical 4D scalar fields must be defined by
$\phi = \sqrt{2{\cal R}_5}\,\hat \phi$ and hence the quartic potential
becomes $\lambda \phi^4/4!$ with $\lambda = \hat \lambda/2 {\cal R}_5
M_5 = \hat  \lambda /N$.

\section{Baryogenesis}

The exact time period for baryogenesis is not known but most
certainly it is completed no later than the electroweak phase
transition.  There are  of course many mechanisms for baryogenesis
which have been proposed in the literature \cite{dol}, too many to
be comprehensive here.  All require baryon number violation, C and
CP violation, and a departure from thermal equilibrium \cite{sak}. 
The original out-of-equilibrium decay (OOED) scenario \cite{ww} is
probably still the simplest of all mechanisms. Originally
formulated in the context of grand unified theories, OOED
involved the decay of a superheavy gauge or Higgs boson with baryon
number violating couplings.  For example, the $X$ gauge boson of
SU(5) couples to both a ($\bar u \bar u$) pair and ($e^- d$).
 The decay rate for $X$ will be  
\be
 \Gamma_{ D}  \simeq   \alpha M_{X}
\ee
  However decays can only begin occurring when the age 
of the Universe is longer
 than the $X$ lifetime   
$\Gamma_D^{-1}$,  i.e., when  $\Gamma_{ D} > $ H  
\be
  \alpha M_{ X}  \ga  N(T)^{ 1/2} T^2/M_{ P} 
\ee
The out-of-equilibrium condition  is 
that at $T = M_{ X},   \Gamma {}_{ D} < H$  
or
 \be
M_{ X} \ga  \alpha M_{ P} (N(M_{ X}))^{ -1/2}  
\sim 10^{18} \alpha {\rm GeV }
\label{mxmin}
\ee
 In this case, we would expect a maximal net baryon asymmetry to be
produced, $n_B/s \sim 10^{-2} \epsilon$, where $\epsilon$ is a measure
of the CP violation in the decay.

In the context of inflation, OOED requires either strong reheating
(or preheating) \cite{local}, or that the decaying particles have
masses less than the inflaton mass so that they can be produced
(necessarily out of equilibrium) by inflaton decays. 
For example, if the inflaton potential carries only a single scale
which is fixed by the magnitude of density fluctuation as measured in
the microwave background radiation, then we can write \cite{cdo}
\be
V(\eta) = \mu^4 ( 1 + O(\eta/M_P)^n + \cdots)
\ee
Typically, 
\be
{\delta \rho \over \rho} \sim O(100) {\mu^2 \over {M_P}^2}
\label{drho}
\ee
so that 
\begin{equation}
{\frac{\mu^2}{M_P^2} = {\rm few} \times{10^{-8}}}
\label{cobemu}
\end{equation}
In this case a relatively light
Higgs is necessary since the inflaton
is typically light ($m_\eta
\sim \mu^2/M_P \sim
O(10^{11})$ GeV)
and the baryon number violating Higgs 
would have to be produced during inflaton decay. 
Clever model building could allow for such a light Higgs, even in the
context of supersymmetry\cite{nt}.  In this case, the baryon asymmetry
is given simply by
\be
{n_B \over s} \sim \epsilon {n_H \over {T_R}^3} 
\sim \epsilon {n_\eta \over {T_R}^3}
\sim \epsilon {T_R \over m_\eta} \sim \epsilon 
\left( {m_\eta \over M_P} \right)^{1/2}
\sim \epsilon {\mu \over M_P}\sim 10^{-4} \epsilon
\ee
where $T_R$ is the reheat temperature after inflation, 
$n_\eta = \rho_\eta/m_\eta 
\sim \Gamma^2{M_P}^2/m_\eta$, and $\Gamma = m_\eta^3 /M_P^2$ is the
inflaton decay rate.

In the context of supersymmetry, there is an extremely natural
mechanism for the generation of the baryon asymmetry utilizing flat
directions of the scalar potential \cite{ad}.  
One can show that
there are many directions in field space such that the scalar
potential vanishes identically
when SUSY is unbroken. That is, 
\be
V = |F|^2 + |D|^2 = 0
\ee
One such example is \cite{ad}
\be
u_3^c = a \qquad s_2^c = a \qquad
-u_1 = v \qquad \mu^- = v \qquad b_1^c = e^{i\phi} \sqrt{v^2 + a^2}
\label{flat}
\ee
where $a,v$ are arbitrary complex vacuum expectation values.
 SUSY breaking lifts this degeneracy so that
\begin{equation}
	V  \simeq \tilde{m}^2 \phi^2
\ee
where $\tilde{m}$ is the SUSY breaking scale and $\phi$ is the direction
 in field space corresponding to the flat direction.
  For large initial values of $\phi$, \ $\phi_o \sim M_{gut}$,
 a large baryon asymmetry can be generated\cite{ad,lin}. This requires
the presence of baryon number violating operators such as $O=qqql$ such that
$\langle O \rangle \neq 0$.  The decay of these
 condensates through such an operator
can lead to a net baryon asymmetry.

When combined with inflation, it is important to verify that the AD
flat directions remain flat. In general, during inflation,
supersymmetry is broken. The gravitino mass is related to the vacuum
energy and $m_{3/2}^2 \sim V/M_P^2 \sim H^2$, thus lifting the flat
directions and potentially preventing the  realization of
the AD scenario as argued in \cite{drt}.
To see this, consider a minimal supergravity model whose
K\"ahler potential is defined by
\be
 G = zz^* + \phi_i^* \phi^i + \ln |\overline{W}(z) + W(\phi)|^2
 \label{min}
\ee
where $z$ is a Polonyi-like field \cite{pol} needed to break
supergravity, and we denote the scalar components of the usual matter
chiral supermultiplets by $\phi^i$. $W$ and $\overline{W}$
are the superpotentials of $\phi^i$ and $z$ respectively. In this
case, the scalar potential becomes
 \be
 V = e^{zz^* + \phi_i^* \phi^i} \left[ |\overline{W}_z +
 z^* (\overline{W} + W)|^2
      + |W_{\phi^i} + \phi^*_i (\overline{W} + W)|^2 - 3|(\overline{W} + W)|^2
    \right]
 \ee
Included in the above expression for $V$, one finds a mass term for
the matter
fields $\phi^i$, $e^G  \phi_i^* \phi^i = m_{3/2}^2 \phi_i^* \phi^i$
\cite{bfs}. As it applies to
all scalar fields (in the matter sector), all flat directions are
lifted by it as well. The above arguments can be generalized to
supergravity models with non-minimal K\"ahler potentials.

There is however a special class of models called no-scale
supergravity models, that were  first introduced in \cite{ns} and have
the remarkable property that the gravitino mass is undetermined at the
tree level despite the fact that supergravity is broken.
No-scale supergravity has been
used heavily in constructing supergravity models in which all mass
scales below the Planck scale are determined radiatively
\cite{nsguts},\cite{nshid}.  
These models emerge naturally in torus \cite{wit} or, for the untwisted sector,
orbifold \cite{orb} compactifications of the heterotic string. 

In no-scale models (or more generally in models which possess a
Heisenberg symmetry \cite{heis}), the K\"ahler potential
becomes
\begin{equation}
G = f(z + z^* - \phi_i^* \phi^i) + \ln |W(\phi)|^2  \label{kpot}
\end{equation}
Now, one can write 
\begin{equation}
V = e^{f(\eta) }\left[
	\left( \frac{f^{\prime 2}}{f^{\prime\prime}} - 3 \right)
		|W|^2
	- \frac{1}{f'} |W_i|^2 \right] 
\label{hpot}
\end{equation}
It is important to notice that the cross term $|\phi_i^* W|^2$ has 
disappeared in the
scalar potential.  Because of the absence of the cross term, flat
directions remain flat even during inflation \cite{gmo}.  
 The no-scale model corresponds to $f = -3 \ln \eta$,
$f^{\prime 2} = 3 f^{\prime\prime}$ and the first term in (\ref{hpot})
vanishes. The potential then takes the form
\be
V = \left[\frac{1}{3} e^{{2 \over 3}f} |W_i|^2 
	\right],
\label{nspot}
\ee
which is positive definite.  The requirement that the vacuum energy vanishes
implies \newline
$\langle W_i\rangle = \langle g_a\rangle = 0 $ at the minimum. As a
consequence $\eta$ is undetermined 
and so is the gravitino mass $m_{3/2}(\eta)$.  

The above argument is only valid at the tree level.
An explicit one-loop calculation~\cite{tr} shows  
that the
effective potential along the flat direction has the form 
\begin{equation}
V_{eff} \sim \frac{g^2}{(4\pi)^2}
\langle V \rangle \left( 
	-2 \phi^2 \log \left(\frac{\Lambda^2}{g^2 \phi^2}\right) 
	+ \phi^2 \right)
	+ {\cal O}(\langle V \rangle)^2 ,
\label{finalV} \end{equation}
where $\Lambda$ is the cutoff of the effective supergravity theory,  
and
has a minimum around $\phi \simeq 0.5 \Lambda$.  
Thus, $\phi_0 \sim M_P$ will be generated and 
in this case the subsequent sfermion oscillations will
dominate the energy density and a baryon asymmetry will result 
which is independent of inflationary parameters as originally  
discussed in
\cite{ad,lin} and will produce $n_B/s \sim O(1)$.
Thus we are left with the problem that the baryon asymmetry in
no-scale type models is too large \cite{eno,gmo,cgmo}. 

In \cite{cgmo}, several possible solutions were presented to dilute
the baryon asymmetry.  These included 1) Moduli decay, 2) the presence
of non-renormalizable interactions, and 3) electroweak effects.
Moduli decay in this context, turns out to be insufficient to bring an
initial asymmetry of order $n_B/s \sim 1$ down to acceptable levels.
However, as a by-product one can show that there is no moduli problem
\cite{pp} either. In contrast, adding non-renormalizable Planck scale
operators of the form $\phi^{2n-2}/M_P^{2n-6}$ leads to a smaller
initial value for $\phi_o$ and hence a smaller value for $n_B/s$.
For dimension 6 operators ($n=4$), a baryon asymmetry of order $n_B/s
\sim 10^{-10}$ is produced. Finally, another possible suppression
mechanism is to  employ the smallness of the
fermion masses.  The baryon asymmetry is  known to be wiped out if the
net $B-L$ asymmetry vanishes because of the sphaleron transitions at
high temperature.   However, Kuzmin, Rubakov and
Shaposhnikov \cite{KRS} pointed out that this  erasure can be
partially circumvented if the individual $(B-3L_{i})$  asymmetries,
where $i=1,2,3$ refers to three generations, do not vanish even when
the total asymmetry vanishes.  Even though there is  still a tendency
that the baryon asymmetry is erased by the chemical equilibrium due to
the sphaleron transitions, the finite mass of the tau lepton shifts
the chemical equilibrium between $B$ and 
$L_{3}$ towards the $B$ side and leaves a finite asymmetry in the 
end.  Their estimate is
\begin{equation}
	B = - \frac{4}{13} \sum_{i} \left(L_{i} - \frac{1}{3}B\right)
		\left( 1 + \frac{1}{\pi^{2}} \frac{m_{l_{i}}^{2}}{T^{2}}\right)
\end{equation}
where the temperature $T \sim T_C \sim 200$~GeV is when the sphaleron 
transition freezes out (similar to the temperature of the electroweak phase
transition)  and 
$m_{\tau}(T)$ is expected to be somewhat smaller than $m_{\tau}(0) = 
1.777$~GeV. Overall, the sphaleron transition suppresses the baryon 
asymmetry by a factor of $\sim 10^{-6}$.  This suppression factor is
sufficient to keep the total baryon asymmetry at a reasonable order of
magnitude in many of the cases discussed above.

Finally, it is necessary to mention one other extremely simple
mechanism based on the OOED of a heavy Majorana neutrino \cite{fy}.
This mechanism
does not require grand unification at all.
By simply adding to the Lagrangian a Dirac and Majorana mass term
 for a new right handed neutrino state, 
\be
{\cal L} \ni M\nu^c\nu^c + \lambda H L \nu^c
\ee
the out-of-equilibrium decays $\nu^c \rightarrow L +  H^*$
 and  $\nu^c \rightarrow L^* + H$ will generate a non-zero 
lepton number $L \neq 0$. The out-out-equilibrium condition
for these decays translates to $10^{-3} \lambda^2 M_P < M$
and $M$ could be as low as $O(10)$ TeV.
(Note that once again in order to 
have a non-vanishing contribution to the C and CP violation
in this process at 1-loop, at least 2 flavors of $\nu^c$ are required.
For the generation of masses of all three neutrino flavors,
3 flavors of $\nu^c$ are required.)
 Sphaleron effects can transfer this lepton asymmetry into a baryon
 asymmetry since now $B - L \neq 0$. A supersymmetric version of 
this scenario
has also been described \cite{cdo,mur}.

{\large\bf Acknowledgments}

As the Inner Space/Outer Space II workshop, and these proceedings
are in memory of Dave Schramm, it is fitting to acknowledge Dave's
role. Dave's research interests were predominantly concerned with the
epoch of Big Bang Nucleosynthesis and the post BBN Universe. His
research interests are known to have been extremely diverse covering
such areas as chemical evolution, cosmic rays, cosmochronology, dark
matter, galaxy formation and mergers, the gamma-ray background,
magnetic fields, mass extinctions, neutrinos, (late-time) phase
transitions, supernovae, ...  In addition, he was a master at
synthesizing and finding relationships between these topics. 
Though pre-BBN cosmology was not Dave's mainstay, he did of course
make important contributions in areas such as baryogenesis and
topological defects.  But Dave's most important contribution was
the early recognition of the impact of cosmology on particle
physics.  He can legitimately be considered a founding father of
astro-particle physics. On a more personal note, it is difficult to
put into words the immense impact that Dave has had on my research, as
an advisor, colleage, and friend.
 This work was supported in part by 
DOE grant DE-FG02-94ER40823 at Minnesota.


\begin{thebibliography}{99}

\bibitem{isos} Inner Space/Outer Space, eds. E.W. Kolb, M.S. Turner,
D. Lindley, K.A. Olive, and D. Seckel, (University of Chicago
Press, Chicago 1985). 

\bibitem{rev} A.D. Linde, {\it Particle
Physics And Inflationary Cosmology} (Harwood, 1990); K.A. Olive,
{\it Phys. Rep.} {\bf 190} (1990) 181; D. Lyth and A. Riotto,
{\it Phys. Rep.} {\bf 314} (1999) 1.

\bibitem{mm} B. Cabrera, {\it Phys. Rev. Lett. } {\bf 48} (1982) 1378.

\bibitem{concl} E.W. Kolb, M.S. Turner,
D. Lindley, K.A. Olive, and D. Seckel, in {\it Inner Space/Outer
Space}, eds. E.W. Kolb, M.S. Turner, D. Lindley, K.A. Olive, and D.
Seckel, (University of Chicago Press, Chicago 1985) p.622.

\bibitem{lyk} J. Lykken, these proceedings.

\bibitem{corrs} C. Lovelace, {\it Phys. Lett.} {\bf B135} (1984) 75;
E.S. Fradkin and A.A. Tseytlin,
{\it Phys. Lett.} {\bf B158} (1985) 316;
C.G. Callan, D. Freidan, E.J. Martinec and M.J. Perry,
{\it Nucl. Phys.}  {\bf B262} (1985) 593;
A. Sen, Phys. Rev. {\bf D32} (1985) 2102;
{\it Phys. Rev. Lett.} {\bf 55} (1985) 1846;
C.G. Callan, I.R. Klebanov and M.J. Perry,
{\it Nucl. Phys.}  {\bf B278} (1986) 78;
S. Deser and N. Redlich,
{\it Phys. Lett.} {\bf B176} (1986) 350;
D. Luest, S. Theisen and G. Zoupanos,
{\it Nucl. Phys.}  {\bf B296} (1988) 800;
J. Lauer, D. Luest and S. Theisen,
{\it Nucl. Phys.}  {\bf B304} (1988) 236;
D. Gross and J. Sloan, {\it Nucl. Phys.} 
{\bf B291} (1987) 41.

\bibitem{w} E. Witten, {\it Nucl. Phys.}  {\bf B471} (1996) 135;
P. Horava, and E. Witten,  {\it Nucl. Phys.}  {\bf
B460} (1996) 506; {\bf B475} (1996) 94;
T. Banks and M. Dine,
{\it Nucl. Phys.} {\bf B479} (1996) 173.

\bibitem{aq} I. Antoniadis and M. Quiros,
{\it Phys. Lett.} {\bf B392} (1997) 61.

\bibitem{dim} I. Antoniadis, {\it Phys. Lett.} {\bf B246} (1990) 377;
I. Antoniadis and K.~Benakli, {\it Phys. Lett.} {\bf B326} (1994) 69;
I. Antoniadis, K.~Benakli and M.~Quiros, {\it Phys. Lett.} {\bf B331}
(1994) 313; 
N. Arkani-Hamed, S. Dimopoulos and G. Dvali,
{\it Phys. Lett.} {\bf B429} (1998) 263; I. Antoniadis, N.
Arkani-Hamed, S. Dimopoulos and G. Dvali, {\it Phys. Lett.} {\bf B436}
(1998) 257; N. Arkani-Hamed, S. Dimopoulos and G. Dvali, {\it Phys.
Rev.} {\bf D59} (1999) 086004;
N. Arkani-Hamed, S. Dimopoulos and J. March-Russell, hep-th/9809124;
K. Dienes, E. Dudas and T. Gherghetta,  {\it Phys. Lett.} {\bf B436}
(1998) 55; {\it Nucl. Phys.} {\bf B537}(1999) 47.


\bibitem{low} 
A. Lukas, B.A. Ovrut and D. Waldram, {\it Phys.Lett.}
{\bf B393} (1997) 65; {\it Nucl. Phys.} {\bf B495} (1997) 365; 
hep-th/9806022; hep-th/9812052;  hep-th/9902071.

\bibitem{nods}C.G. Callan, D. Friedan, E.J. Martinec and
 M.J. Perry, {\it Nucl. Phys.} {\bf B262} (1985) 593;
 C.G. Callan, I.R. Klebanov and M.J. Perry, {\it Nucl. Phys.}
{\bf B278} (1986) 78; D.G. Boulware and S. Deser, {\it Phys. Lett.}
{\bf B175} (1986) 409; S. Kalara, C. Kounnas and K.A. Olive, {\it
Phys. Lett.} {\bf B215} (1988) 265;
S. Kalara and K.A. Olive, {\it Phys. Lett.} {\bf
B218} (1989) 148;
M.C. Bento and O. Bertolami, {\it Phys. Lett.}
{\bf B368} (1996) 198.

\bibitem{clo} B.A. Campbell, A. Linde and K.A. Olive,
{\it Nucl. Phys.} {\bf B355} (1991) 146.

\bibitem{lindil} R.C. Myers, Phys. Lett. {\bf B199} (1987) 371;
M. Mueller, Nucl. Phys. {\bf B337} (1990) 37;
I. Antoniadis, C. Bachas, J. Ellis and
D.V. Nanopoulos, Phys. Lett. {\bf B211}
(1988) 393; Nucl. Phys. {\bf B328} (1989) 117.


\bibitem{pbb} G. Veneziano, {\it Phys. Lett.} {\bf 265} (1991) 287;
M. Gasperini and G. Veneziano, {\it Astropart. Phys.} {\bf 1} (1993) 317.

\bibitem{ge}  R. Brustein and G. Veneziano, Phys. Lett. {\bf B329} (1994)
429; N. Kaloper, R. Madden and K.A. Olive, Nucl. Phys.
{\bf B452} (1995) 677; E.J. Copeland, A. Lahiri and D. Wands, Phys. Rev. {\bf D50}
(1994) 4868; Phys. Rev. {\bf D51} (1995) 1569;  N. Kaloper, R. Madden and
K.A. Olive, Phys. Lett. {\bf B371} (1996) 34; R. Easther, K. Maeda and D.
Wands, Phys. Rev. {\bf D53} (1996) 4247.

\bibitem{bm} R. Brustein and R. Madden,
{\it Phys. Lett.} {\bf B410} (1997) 110; {\it Phys. Rev.} {\bf
D57} (1998) 7121998; A. Ghosh, R. Madden, and G. Veneziano,
hep-th/9908024.


\bibitem{ikko} N. Kaloper, I. Kogan and K.A. Olive,
{\it Phys. Rev.} {\bf D57} (1998) 7340.

\bibitem{linf} D. Lyth, {\it Phys. Lett.} {\bf B448} (1999) 191;
N. Kaloper and A. Linde, hep-th/9811141; N. Arkani-Hamed, S.
Dimopoulos, N. Kaloper, and J. March-Russell, hep-th/9903224.


\bibitem{ekoy} J. Ellis, N. Kaloper, K.A. Olive and J. Yokoyama, 
{\it Phys. Rev.} {\bf D59} (1999) 103503.

\bibitem{star} A.A. Starobinsky, {\it Sov. Astron. Lett.} {\bf 9} (1983)
302.

\bibitem{star2} A.A. Starobinsky,
{\it Sov. Astron. Lett.} {\bf 9} (1983) 302.

\bibitem{whitt} B. Whitt, {\it Phys. Lett.} {\bf 145B} (1984) 176.

\bibitem{chaotic} A.D. Linde, {\it Phys. Lett.} {\bf B129} (1983)
181.

\bibitem{corr} M.S. Duff, {\it Nucl. Phys.} {\bf B125} (1977) 334; P.C.W.
Davies, S.A. Fulling, S.M. Christensen and T.S. Bunch, {\it Ann.
Phys.} {\bf 109} (1977) 108; T.S. Bunch and P.C.W. Davies,
{\it Proc. R. Soc. Lond.} {\bf A356} (1977) 569.

\bibitem{dc} J.S. Dowker and R. Critchley, {\it Phys. Rev.}
{\bf D13} (1976) 3224.

\bibitem{vil} A. Vilenkin, {\it Phys. Rev.} {\bf D32} (1985) 2511.

\bibitem{fp} R. Fabbri and M.D. Pollock, {\it Phys. Lett.} {\bf 125B} (1983)
445.


\bibitem{Maeda} K. Maeda, {\it Phys. Rev.} {\bf D39} (1989) 3159.


\bibitem{topological} A.D. Linde, {\it Phys. Lett.} {\bf B327} (1994)
208; A. Vilenkin, {\it Phys. Rev. Lett.} {\bf 72} (1994) 3137.





\bibitem{yko1} P. Kanti and K.A. Olive, {\it Phys. Rev.} {\bf  D60}
(1999) 043502.

\bibitem{yko2} P. Kanti and K.A. Olive, {\it Phys. Lett. B} (1999) in 
press, hep-ph/9906331.

\bibitem{liddle} A.R. Liddle, A. Mazumdar and F.E. Schunck, {\it Phys.
Rev.} {\bf D58} (1998) 061301.

\bibitem{mw} K.A. Malik and D. Wands, {\it Phys. Rev.} {\bf  D59}
(1999) 123501.


\bibitem{em} K. Enqvist and J. Maalampi, {\it Phys. Lett.} {\bf B180}
(1986) 14.


\bibitem{maz} E.J. Copeland, A. Mazumdar, and N.J. Nunes,
astro-ph/9904309.


\bibitem{dol}for reviews see: E.W. Kolb and M.S. Turner, {\it Ann.
Rev. Nucl.Part. Sci.} {\bf 33} (1983) 645;
A. Dolgov, {\it Phys. Rep.} {\bf 222} (1992) 309.


\bibitem{sak} A.D. Sakharov, {\it JETP Lett.} {\bf 5} (1967) 24.


\bibitem{ww} S. Weinberg, {\it Phys. Rev. Lett.} {\bf 42}, (1979) 850;
	D. Toussaint, S. B. Treiman, F. Wilczek, and A. Zee, {\it Phys. Rev.}
 {\bf D19} (1979) 1036;
J.N. Fry, K.A. Olive, and M.S. Turner, {\it Phys. Rev.}
{\bf D22} (1980) 2953; {\bf D22} (1980) 2977; {\it Phys. Rev. Lett.}
{\bf 45} (1980) 2074;
E.W. Kolb and S. Wolfram, {\it Phys. Lett.} {\bf B91}
(1980) 217; {\it Nucl. Phys.} {\bf B172} (1980) 224.

\bibitem{local} D. J. H. Chung, E. W. Kolb, A. Riotto,
{\it Phys. Rev.} {\bf D60} (1999) 063504.

\bibitem{cdo}B. Campbell, S. Davidson, and K.A. Olive, 
{\it Nucl.Phys.} {\bf B399}
(1993) 111.

\bibitem{nt} D.V. Nanopoulos and K. Tamvakis, {\it Phys. Lett.}
{\bf B114} (1982) 235.

\bibitem{ad}I. Affleck and M. Dine, {\it Nucl. Phys.} {\bf B249} (1985)
361.
\bibitem{lin} A.D. Linde, {\it Phys. Lett.} {\bf B160} (1985) 243.

\bibitem{drt} M.Dine, L. Randall, and S. Thomas,
{\it Phys. Rev. Lett.} {\bf 75} (1995) 398; {\it Nucl.Phys.}
{\bf B458} (1996) 291.

\bibitem{pol} J. Polonyi, Budapest preprint KFKI-1977-93.
\bibitem{bfs} R. Barbieri, S. Ferrara, and C.A. Savoy, Phys. Lett. {\bf B119}
(1982) 343.

\bibitem{ns} E. Cremmer, S. Ferrara, C. Kounnas and D.V. Nanopoulos, Phys. Lett.
    {\bf 133B} (1983) 61.
\bibitem{nsguts} J. Ellis, A.B. Lahanas, D.V. Nanopoulos and K.
     Tamvakis, Phys. Lett. {\bf 134B} (1984) 429;
J. Ellis, C. Kounnas and D.V. Nanopoulos, Nucl. Phys.
    {\bf B241} (1984) 406; J. Ellis, C. Kounnas and D.V. Nanopoulos, Nucl. Phys.
     {\bf B247} (1984) 373; for a review see: A.B. Lahanas 
and D.V. Nanopoulos, Phys. Rep. {\bf 145} (1987) 1.
\bibitem{nshid} J.D. Breit, B. Ovrut, G. Segr\'e, Phys.\ Lett.\ {\bf 162B} 
(1985) 303;
P.\ Bin\'etruy and M.K. Gaillard, Phys.\ Lett.\ {\bf 168B} (1986)  347;
P.\ Bin\'etruy, S.\ Dawson M.K. Gaillard and I.\ Hinchliffe, 
Phys.\ Rev.\ {\bf D37} (1988) 2633 and references therein.
\bibitem{wit} E. Witten, {\it Phys.\ Lett.} {\bf 155B} (1985) 151.
\bibitem{orb} S. Ferrara, C. Kounnas and M. Porrati, {\it Phys. Lett.} 
{\bf 181B} (1986) 263;
L.J. Dixon, V.S. Kaplunovsky and J. Louis, {\it Nucl. Phys.} 
{\bf B329} (1990) 27; S. Ferrara, D. L\"ust, and S. Theisen, 
{\it Phys. Lett.} {\bf 233B} (1989) 147.

\bibitem{heis} P.~Binetruy and M.K.~Gaillard, Phys. Lett. {\bf B195}
(1987) 382.


\bibitem{gmo} M.K. Gaillard, H. Murayama, and K.A. Olive, Phys. Lett. 
{\bf B355} (1995) 71.

\bibitem{tr} M.K.~Gaillard and V.~Jain, Phys. Rev. {\bf D49}
(1994) 1951;
M.K.~Gaillard, V.~Jain and K. Saririan, Phys. Lett. {\bf B387}
 (1996) 520 and Phys. Rev. {\bf D55}  (1997) 833.


\bibitem{eno} J. Ellis, D.V. Nanopoulos, and K.A. Olive,  
Phys. Lett. {\bf  B184} (1987) 37.

\bibitem{cgmo} B.A. Campbell, M.K. Gaillard, H. Murayama, and K.A.
Olive, {\it Nucl. Phys.} {\bf B538} (1999) 351. 

\bibitem{pp} G.D. Coughlan, W. Fischler, E.W. Kolb, S. Raby and G..G.
 Ross, Phys. Lett. {\bf B131} (1983) 59.


\bibitem{KRS} V.A. Kuzmin, V.A. Rubakov, and M.E. Shaposhnikov,  Phys. 
Lett. {\bf 191B} 171 (1987); see also, H. Dreiner and G. Ross, Nucl. Phys.
{\bf B410} (1993) 188; S. Davidson, K. Kainulainen, and K.A. Olive, Phys. Lett.  
{\bf B335} (1994) 339. 


\bibitem{fy} M. Fukugita and T. Yanagida, {\it Phys. Lett.} 
{\bf B174} (1986) 45.

\bibitem{mur} H. Murayama, H. Suzuki, T. Yanagida, and J. Yokoyama,
{\it Phys. Rev. Lett.} {\bf 70} (1993) 1912.


\end{thebibliography}
\end{document}